\documentclass[conference]{IEEEtran}
\IEEEoverridecommandlockouts

\usepackage[linesnumbered]{algorithm2e}
\usepackage{enumitem}
\usepackage{color}
\usepackage{theorem}
\usepackage{times,amsmath,epsfig}
\usepackage{amssymb}
\usepackage{cite}
\usepackage{multirow}

\usepackage{hyperref}
\usepackage{algorithmic}
\usepackage{rotating}
\usepackage{cleveref}

\usepackage[utf8]{inputenc}
\usepackage{color}
\usepackage{theorem}
\usepackage{times,amsmath,epsfig}
\usepackage{multirow}
\usepackage{amsfonts}
\usepackage{spalign}
\usepackage[linesnumbered]{algorithm2e}

\usepackage{graphicx}

\setlist[itemize]{leftmargin=*}

\input{mysymbol.sty}

\title{Learning the Topology of a Simplicial Complex Using Simplicial Signals: A Greedy Approach}
\author{Andrei Buciulea, Elvin Isufi, Geert Leus, and Antonio G. Marques\thanks{Work supported by the EU H2020 Grant Tailor (No 952215, agreements 31 and 82); the Dutch Grant GraSPA (No 19497) financed by the Netherlands Organization for Scientific Research (NWO); the Spanish  (MCIN/AEI/10.13039/501100011033) Grants PID2019-105032GB-I00, TED2021-130347B-I00 \& PID2022-136887NB-I00; and by the Autonomous Community of 
Madrid within the ELLIS Unit Madrid framework, and URJC/CAM grant F861.~~A. Buciulea \& A.G. Marques are with King Juan Carlos University, Madrid, Spain (e-mail:{andrei.buciulea, antonio.garcia.marques}@urjc.es).     
E. Isufi \& G. Leus are with the Delft University of Technology, Delft, The Netherlands (e-mail:{e.isufi-1, g.j.t.leus}@tudelft.nl).
}}

\input{mysymbol.sty}

\def \ccalEobs {\ccalE^{\ccalO}}

\usepackage{mathtools}
\DeclarePairedDelimiterX{\norm}[1]{\lVert}{\rVert}{#1}

\newtheorem{mylemma}{\bf Lemma}

\begin{document}

\maketitle

\begin{abstract} 
Graphs are ubiquitous to model the irregular (non-Euclidean) structure of complex data, but they are limited to pairwise relationships and fail to model the complexities of the datasets exhibiting higher-order interactions. In that context, simplicial complexes (SCs) are emerging as a tractable candidate to handle such domains. The first step in using SC-based processing and learning schemes is to identify the topology of the SC, which is the problem investigated in this paper. In particular, we assume that we observe a number of signals (features) associated with the nodes of the SC (simplices of order 0) as well as signals (features) associated with a subset of the edges of the SC (simplices of order 1). The goal is then to use these signals to learn the remaining edges as well as the triangles that are filled (simplices of order 2). To address this problem, we assume that the signals are smooth on the unknown SC and that the higher-order relations are sparse. We then postulate the learning problem as a nonconvex optimization and develop an efficient (block-coordinate) algorithm to identify the SC topology.

\end{abstract}

\section{Introduction}

Inferring the support of complex data such as graph-based data is a long-standing research challenge in statistics, signal processing (SP), and machine learning \cite{friedman2008sparse,timme2007revealing,mateos2019connecting}. This is motivated by the fact that graphs are a versatile tool that can exploit intrinsic dependencies in multivariate and coupled dynamics. However, graphs only capture pair-wise interactions, failing to capture higher-order dependencies (three-wise and so on) that are key to, e.g., reveal social circles \cite{majhi2022dynamics,bick2023higher} or model dependencies between data defined over tuples such as flows \cite{jia2019graph,money2022online,liu2023unrolling}. This work studies the problem of inferring higher-order dependencies from noisy and partially observed data.

Hypergraphs are often considered to represent higher-order data structures but, because of their high degrees of freedom and lack of unique representation, inferring hyper-structures is difficult as it is challenging to impose reasonable assumptions on the data behavior \cite{young2021hypergraph,tang2023learning}. Topological structures such as simplicial complexes (SCs) represent a viable alternative as they impose a hierarchical structure within that data that can facilitate processing \cite{barbarossa2020topological,schaub2021SPOnHigherOrder,yang2022simplicial}, allow for an algebraic and spectral representation \cite{grady2010discrete}, and respect realistic data priors such as divergence-free and curl-free that are common in flow-data. Some recent works have been proposed to infer such topological structures from data \cite{barbarossa2020topological,buciulea2024learning,wang2022full}. Specifically, \cite{barbarossa2020topological} extends graph SP techniques to infer three-wise interactions between flow data, while in our recent work \cite{buciulea2024learning} we propose a Volterra model to jointly infer graphs and simplicial structures only from nodal observations. 
Finally, the work in \cite{wang2022full} proposes a statistical model to infer an SC structure from binary interactions.

Differently from these works, we infer here the structure of a second-order SC when (partial and noisy) observations are obtained on the nodes and edges. From a geometric and SP perspective, we consider available all node signals as well as a collection of edge signals at a subset of edges, and we aim to infer both the remaining edges and the triangle connections. Towards this aim, we consider conventional sparsity assumptions on the underlying structure as well as smoothness on the node and edge signal values, ultimately, respecting SC signal behavior encountered in real-world applications. The considered formulation gives rise to a challenging highly nonconvex problem. We handle this by proposing a suitable relaxation followed by a coordinate descent algorithm. Our novel approach leads to a 3-step algorithm where edges and triangles are identified using a greedy algorithm.


\section{Notation and Mathematical Preliminaries}

This section introduces the preliminary concepts about SCs that will be used in the remainder of the paper.


\vspace{1mm}
\noindent \textbf{SCs and incidence matrices:} Consider the sets $\ccalN$, $\ccalE$ and $\ccalT$, where $\ccalN$ collects $N=|\ccalN|$ nodes,  $\ccalE \subset \ccalN \times \ccalN$ collects the $E=|\ccalE|$ active edges connecting those nodes, and $\ccalT \subset \ccalN \times \ccalN \times \ccalN$ collects $T = |\ccalT|$ filled triangles relating triplets of nodes.  For a triplet $(\ccalN, \ccalE, \ccalT)$ qualifying as an SC of order 2, it must hold that if $(v_1,v_2,v_3)\in\ccalT$ then the three edges $(v_1,v_2)$, $(v_1,v_3)$, and $(v_2,v_3)$, must be elements of $\ccalE$. The elements of $\ccalN$, $\ccalE$ and $\ccalT$ are also referred to as simplices of order 0, 1, and 2, respectively. If the SC is undirected, the maximum value for $E$ and $T$ is $\bar{E}= \binom{N}{2}$ and $\bar{T}= \binom{N}{3}$, respectively.

As in graphs, the topology of an SC is oftentimes represented by Laplacian-based matrices that are briefly discussed next. First, we discuss the incidence matrices, where we define some reference orientation for each simplex to ease computations. Specifically, let $\bbB_1 \in \reals^{N \times E}$ and $\bbB_2 \in \reals^{E \times T}$ be the incidence matrices that describe the membership relation in our SC of order two~\cite{schaub2021SPOnHigherOrder}.
\begin{itemize}
    \item Each column of $\bbB_1$ corresponds to an active edge. All entries of a column are zero except for the two identifying the elements (nodes) that participate in the link, which are either $+1$ or $-1$ based on a fixed ordering of the vertices that define the edge orientations. 
    %
	\item Similarly, each column of $\bbB_2$ corresponds to an active (filled) triangle. 
    All entries of a column of $\bbB_2$ are zero, except for the three identifying the edges that participate in the triangle, which are again either a $+1$ or a $-1$ based on the edges being aligned or not with the triangle orientation. 
\end{itemize}
Based on the oriented versions of $\bbB_1$ and $\bbB_2$, we form the matrices 
%
%
$\bbL_0 = \bbB_1\bbB_1^\top$ and $\bbL_1 = \bbB_1^\top\bbB_1 + \bbB_2\bbB_2^\top$, 
which correspond to the classical combinatorial node Laplacian and the so-called Hodge Laplacian, respectively. When dealing with the incidence matrices, the simplicial closure constraint (SCC) requiring that triangles can only be filled if the 3 associated edges are present holds true if and only if 
\begin{equation}\label{E:simp_clos_constr}
	\bbB_1\bbB_2=\bb0.
\end{equation} 
Since the latter implies that the range spaces of $\bbB_1^\top$ and $\bbB_2$ are orthogonal, it readily follows that    $\reals^E=\mathrm{span}(\bbB_1^\top)\oplus\mathrm{span}(\bbB_2)\oplus\mathrm{kernel}(\bbL_1)$ for any SC. 

\vspace{1mm}
\noindent \textbf{SC signals:} In our setup, we consider that signals are attached to nodes and edges of the SC. To that end, we denote as $\bbX_0 = [\bbx_1^0,...,\bbx_{P_0}^0] \in \mathbb{R}^{N \times P_0}$ the matrix collecting all the nodal observations ($P_0$ per node). Additionally, we consider that a subset of edges $\ccalEobs\subset\ccalE$ with cardinality $E^{\ccalO} < E$ is known and that we observe their edge/flow signals.
Then, we denote as $\bbX_1^{\ccalO} = [\bbx_1^{\ccalO, 1},...,\bbx_{P_1}^{\ccalO, 1}] \in \mathbb{R}^{E^{\ccalO} \times P_1}$ the matrix collecting the edge signal observations. 
The matrix of all edge signals is given by $\bbX_1 = [\bbx_1^{1},...,\bbx_{P_1}^{1}] \in \mathbb{R}^{E \times P_1}$.


\vspace{1mm}
\noindent \textbf{Smoothness:} Nodal observations are considered smooth if they do not differ much over the edges. This is typically quantified via the Laplacian quadratic form
\[ \text{trace} (  \bbX_0^\top   \bbL_0  \bbX_0 ) =  \text{trace} (  \bbX_0 \bbX_0^\top \bbB_1\bbB_1^\top ).  \]
It can be shown that the smaller this measure, the more smooth the nodal observations are over the edges \cite{mateos2019connecting}.

Next, the earlier expression $\reals^E=\mathrm{span}(\bbB_1^\top)\oplus\mathrm{span}(\bbB_2)\oplus\mathrm{kernel}(\bbL_1)$ allows us to decompose any edge signal $\bbx^1$ as $\bbx^1 = \bbB_1^\top \bbv + \bbB_2 \bbt + \bbx^1_{\rm h}$.
The terms of this so-called Hodge decomposition can be interpreted as follows.
\begin{itemize}
\item The term $\bbB_1^\top \bbv$ is called the gradient flow and is induced by the difference of the vertex signal $\bbv$ over the edges. Related to this we can define the divergence 
operator $\bbB_1 \bbx^1$ which measures the divergence of an edge
flow. The $i$th element corresponds to the flow passing through the $i$th node. If $\bbB_1 \bbx^1 = \bb0$ the flow is divergence-free.
\item On the other hand, $\bbB_2 \bbt$ is called the curl flow and consists of the cyclic flow along the edges of all active triangles induced by the triangle signal $\bbt$. Related to this, we can define the curl operator $\bbB_2^\top \bbx^1$ which yields a triangle signal that measures the curl of an edge flow. The $i$th element corresponds to the sum of the flows of each edge forming the $i$th triangle. If $\bbB_2^\top \bbx^1 = \bb0$ the flow is curl-free.
\item The remaining term $\bbx^1_{\rm h}$ is called the harmonic flow. 
\end{itemize}
Real flow signals often have a small divergence and/or small curl. In this paper, we particularly assume that flows have a small curl since that can provide us with some information on $\bbB_2$. Specifically, assuming that the edge signal $\bbX_1$ has a small curl (or is ``smooth'' over the triangles) means that the following measure should be small \cite{barbarossa2020topological,yang2022simplicial}: 
\[    \|  \bbB_2^\top \bbX_1 \|_F^2  = \text{trace} (  \bbX_1 \bbX_1^\top \bbB_2\bbB_2^\top ). \]


\section{Problem Statement}

We start by formalizing our SC-learning problem.

\noindent \textbf{Problem 1} \emph{Given a set of nodes $\ccalN$, a collection of (noisy) node-signal observations $\bbX_0$, and a collection of (noisy) edge-signals $\bbX_1^\ccalO$ observed at some edges ($\ccalEobs = \ccalE/ \ccalE^{\ccalU}$) of an SC of order 2, find the sets $\ccalE$ and $\ccalT$ that, together with $\ccalN$, define the SC, under assumptions (AS1-AS4): (AS1) The number of edges is small; (AS2) The number of filled triangles is small; (AS3) The node-observations $\bbX_0$ are smooth on the underlying graph \cite{DongLaplacianLearning,egilmez2016graph,mateos2019connecting}; and (AS4) The edge-observations $\bbX_1$ have a low curl (or are ``smooth'' over the triangles).}






Taking into account Assumptions (AS1-AS4), a nonlinear optimization formulation of the SC-learning problem that promotes sparsity and signal smoothness, guarantees the SC structure, and accounts for the observed edge labels is
\begin{subequations}\label{E:SC_OPT_v0}
\begin{alignat}{2}%
	\!\!&\! \min_{\bbX_1,\{\bbB_i\}_{i=1}^2} &&   \|\bbB_1\bbB_1^\top\|_0 + \|\bbB_2\bbB_2^\top\|_0+\tr(\bbX_0\bbX_0^\top\bbB_1\bbB_1^\top)  \nonumber\\
	\!\!&\! &&+ \tr(\bbX_1\bbX_1^\top\bbB_2\bbB_2^\top)+\|\bbTheta\bbX_1-\bbX_1^\ccalO\|_F^2    \label{E:SC_OPT_v0_obj}\\
	\!\!&\! \hspace{4.5mm}\mathrm{\;\;s. \;t. } && 
	\bbB_1\in \ccalB_1 ,\;\bbB_2\in \ccalB_2, \label{E:SC_OPT_v0_c1}\\
	\!\!&\! && \bbB_1\bbB_2 = \bb0, \label{E:SC_OPT_v0_c2}\\
	\!\!&\! && [\bbB_1\bbB_1^\top]_{ij}=-1 \;\;\text{forall} \;(i,j)\in \ccalEobs, \;\; \label{E:SC_OPT_v0_c3}
\end{alignat} 
\end{subequations}
where $\bbTheta\in\{0,1\}^{E^{\ccalO}\times E}$ is the appropriate edge sampling matrix, and $\ccalB_1$, $\ccalB_2$ denote the sets of feasible incidence matrices. The first two terms in the objective account for the sparsity in (AS1) and (AS2), the third and fourth ones for the smoothness in (AS3) and (AS4), and the last one for the noise in $\bbX_1$. Notice that, to help readability, regularizer weights have been omitted. Constraints \eqref{E:SC_OPT_v0_c1}-\eqref{E:SC_OPT_v0_c2} guarantee that we obtain a valid SC and \eqref{E:SC_OPT_v0_c3} accounts for the observed edges. The optimization: i) requires defining the auxiliary variable $\bbX_1$; and ii) the constraints in \eqref{E:SC_OPT_v0_c1} include a minimum number of edges and triangles that prevent the (trivial) all-zero solution. 


Formulation~\eqref{E:SC_OPT_v0} has three main challenges: 1) It lacks convexity due to the $\ell_0$ norms and the multilinear terms; 2) Enforcing that $\bbB_1$ and $\bbB_2$ are feasible incidence matrices is challenging by construction; 
%
and 3) Guaranteeing that the solution satisfies the SCC is nontrivial. 
Sec. \ref{S:Greedy_SCL_formulation} introduces a tractable approach to mitigate some of these challenges.


\subsection{SC-learning approach via simplex selection}\label{S:Greedy_SCL_formulation}

Our two key ideas to render the optimization in \eqref{E:SC_OPT_v0_obj} more tractable are: K1) recast the optimization over the incidence/Laplacian matrices as an edge/triangle selection problem and K2) reformulating the SCC.

To describe idea K1), consider the \emph{complete} SC $(\ccalN,\bar{\ccalE},\bar{\ccalT})$, where $\bar{\ccalE}=\ccalN\times\ccalN$ and $\bar{\ccalT}=\ccalN\times\ccalN\times\ccalN$. We denote the incidence matrices of the complete SC $(\ccalN,\bar{\ccalE},\bar{\ccalT})$ as $\bar{\bbB}_1\in\reals^{N\times\bar{E}}$ and $\bar{\bbB}_2\in\reals^{\bar{E}\times\bar{T}}$. Let us consider now a generic (non-complete) SC $(\ccalN,\ccalE,\ccalT)$ and associate two binary vectors ($\bbw_1\in\{0,1\}^{\bar{E}}$ and $\bbw_2\in\{0,1\}^{\bar{T}}$) with it. An entry of $\bbw_1$ is 1 iff the corresponding column of $\bar{\bbB}_1$ identifies an edge present in $\ccalE$. Similarly, an entry of $\bbw_2$ is 1 iff the corresponding column of $\bar{\bbB}_2$ identifies a triangle present in $\ccalT$. 
By reformulating the optimization in terms of the selection vectors $\{\bbw_i\}_{i=1}^2$ we have that a sparse SC boils down to imposing sparsity on $\{\bbw_i\}_{i=1}^2$. Secifically, the quadratic terms $\bbB_i \bbB_i^\top$ can be written as $\bar{\bbB}_i\diag(\bbw_i)^2\bar{\bbB}_i^\top=\bar{\bbB}_i\diag(\bbw_i)\bar{\bbB}_i^\top$, which are linear in $\bbw_i$. Note that as a consequence of K1) we also need to introduce the full edge signal matrix $\bar{\bbX}_1 \in \mathbb{R}^{\bar{E} \times P_1}$.


To describe idea K2), recall that a valid SC must satisfy the SCC. That is, a triangle $(i,j,k)$ can belong to $\ccalT$ only if the 3 edges of the triangle belong to the edge set $\ccalE$. To formalize this, let us consider the non-oriented edge-to-triangle incidence matrix associated with the complete SC. Mathematically, let $\bar{\bbB}_2^+\in\{0,1\}^{\bar{E}\times\bar{T}}$ be a binary matrix so that $[\bar{\bbB}_2^+]_{l,t}=1$ implies that the edge indexed by $l$ is one of the tree links involved in the triangle indexed by $t$. With this notation at hand, the pair $(\bbw_1,\bbw_2)$ forms a valid SC (i.e., satisfies the SCC) if and only if $(\bbone -\bbw_1)^\top \bar{\bbB}_2^+ \bbw_2=0$.    

With all this in place, we reformulate problem \eqref{E:SC_OPT_v0} as 
\begin{subequations}\label{E:SC_OPT_v1}
\begin{alignat}{2}
	\!\!&\!\min_{\bar{\bbX}_1, \{\bbw_i\}_{i=1}^2} &&\!\! \!\!  \alpha_1\|\bbw_1\|_0+\alpha_2\|\bbw_2\|_0 +  \beta_1\tr(\bbX_0\bbX_0^\top\bar{\bbB}_1\diag(\bbw_1)\bar{\bbB}_1^\top)   \nonumber \\
	\!\!&\hspace{9mm} +&&\!\!\!\!  \beta_2\tr(\bar{\bbX}_1 \bar{\bbX}_1^\top\!\bar{\bbB}_2\diag(\bbw_2)\bar{\bbB}_2^\top) \!+\!\| \bar{\bbTheta} \bar{\bbX}_1\!-\!\!\bbX_1^\ccalO\|_F^2  \label{E:SC_OPT_v1_obj} \\ 
	\!\!&\!\hspace{4.5mm} \mathrm{\;\;s. \;t. } && \bbw_1 \in \{0,1\}^{\bar{E}},  \;\bbw_2 \in \{0,1\}^{\bar{T}}, \label{E:SC_OPT_v1_c1}\\
	\!\!&\! && 	(\bbone -\bbw_1)^\top \bar{\bbB}_2^+ \bbw_2=0, \label{E:SC_OPT_v1_c2}\\
\!\!&\! && [\bbw_1]_{l}=1 \;\;\text{forall} \;l\in \ccalEobs,\;\; \label{E:SC_OPT_v1_c3}\\  
	\!\!&\! &&   \|\bbw_1\|_0 \geq E^{\min} \;  \text{ and } \; \|\bbw_2\|_0 \geq T^{\min}. \label{E:SC_OPT_v1_c5}
\end{alignat} 
\end{subequations}
Here $\bar{\bbTheta}\in\{0,1\}^{ E^{\ccalO}\times \bar{E}}$ is the edge sampling matrix that selects the observed edges from all possible edges.
Note that \eqref{E:SC_OPT_v1_obj}-\eqref{E:SC_OPT_v1_c3} are the counterpart to \eqref{E:SC_OPT_v0_obj}-\eqref{E:SC_OPT_v0_c3}, while \eqref{E:SC_OPT_v1_c5} is introduced to prevent the trivial all-zero solution. 
The updated formulation is more tractable than \eqref{E:SC_OPT_v0} since: 
a) the number of variables is smaller; 
b) the number of constraints is much smaller (e.g., \eqref{E:SC_OPT_v0_c2} is a matrix constraint while \eqref{E:SC_OPT_v1_c2} is a scalar constraint);  
c) the optimization variables belong to sets that are easier to describe; and
d) several of the multilinear terms are now linear. 
%
However, the optimization in \eqref{E:SC_OPT_v1_obj}-\eqref{E:SC_OPT_v1_c5} is still challenging due to $\{\bbw_i\}_{i=1}^2$ being binary and the two bilinear terms (one in the SCC constraint and another one in the smooth edge term). Sec. \ref{S:GredyAlgorithm} discusses an efficient algorithm to deal with this.

\begin{figure*}[ht]
 \centering	 \includegraphics[width=1\textwidth]{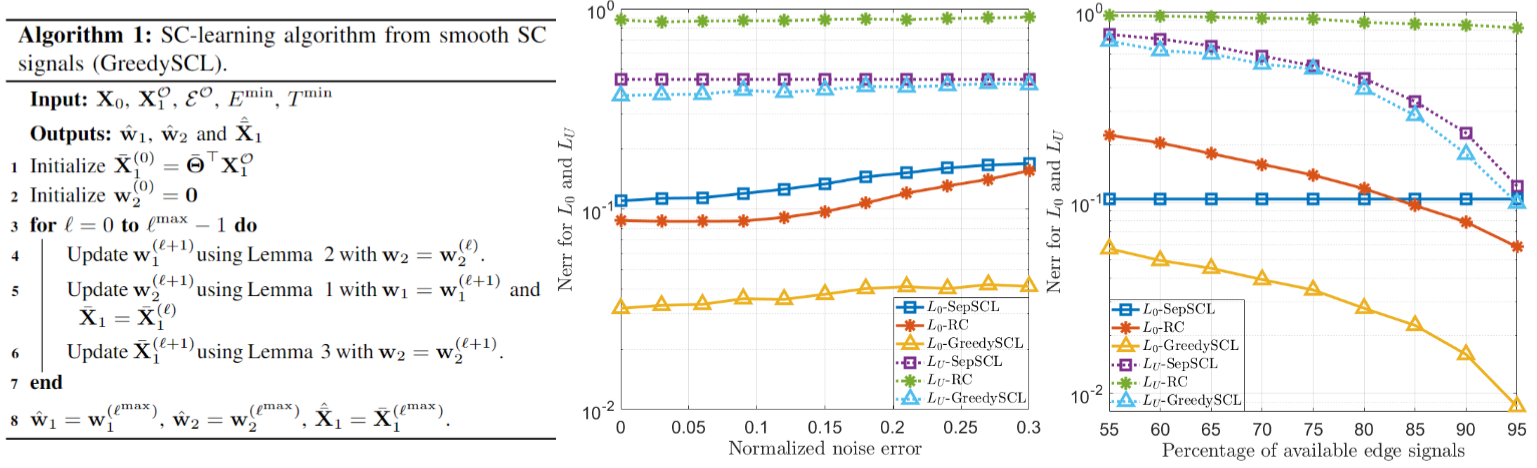}
          \vspace{-0.6cm}
	\caption{(a) Algorithm for joint graph \& SC learning from smooth node \& edge signals. Estimation error for $\bbL_0$ \& $\bbL_U$ for different SC-learning methods varying (b) the noise level in the node signals and (c) the percentage of observed edge signals.}	\label{F:exp_1}
\end{figure*}

\section{Algorithmic approach: Greedy SC-Learning}\label{S:GredyAlgorithm}

Our approach to address the optimization is to: i) remove constraint ($\bbone -\bbw_1)^\top \bar{\bbB}_2^+ \bbw_2=0$ [cf. \eqref{E:SC_OPT_v1_c2}] and augment the objective with $\gamma(\bbone-\bbw_1)^\top \bar{\bbB}_2^+ \bbw_2$; ii) implement an alternating optimization approach to deal with the bilinearities; and iii) implement an (optimal) greedy optimization to deal with sparsity and the binary constraints. Regarding i), note that all terms in \eqref{E:SC_OPT_v1_c2} are nonnegative and, as a result, the constraint acts as a penalty without the need of using any norm. Regarding ii), alternating approaches are particularly suitable for bilinear and biconvex problems \cite{gorski2007biconvex}. Overall, our approach results in a 3-step iterative algorithm that optimizes over one block of variables while keeping the other two constant.

We analyze each of the three subproblems next. To facilitate exposition, we first present the optimization for $\bbw_2$, then for $\bbw_1$ and finally the optimization over $\bar{\bbX}_1$. 

\vspace{.2cm}
\noindent \emph{A. Optimization of the triangle selection vector:} We consider $\bbw_1$ and $\bar{\bbX}_1$ given and optimize $\bbw_2$. The formal problem is
\begin{alignat}{2}
	\!\!&\!\min_{\bbw_2} \alpha_2\|\bbw_2\|_0+ \beta_2\tr(\bar{\bbX}_1 \bar{\bbX}_1^\top\bar{\bbB}_2\diag(\bbw_2)\bar{\bbB}_2^\top)  
	\label{E:SC_OPT_v1_subp2_c1e} \\ 
	\!\!&\! + \gamma (\bbone\!-\!\bbw_1)^\top \bar{\bbB}_2^+ \bbw_2~\mathrm{\;\;s. \;t. } \, \bbw_2 \!\in\! \{0,1\}^{\bar{T}}\!\!,  \|\bbw_2\|_0 \!\geq T^{\min}\! . \nonumber
\end{alignat} 

Let us write the second  and third terms in the objective as
\begin{align}
	 &\;\tr(\bar{\bbX}_1 \bar{\bbX}_1^\top\bar{\bbB}_2\diag(\bbw_2)\bar{\bbB}_2^\top) \! = \! \textstyle \sum_{t=1}^{\bar{T}}[\bar{\bbB}_2^\top \bar{\bbX}_1 \bar{\bbX}_1^\top\bar{\bbB}_1]_{tt}[\bbw_2]_t, \notag \\
	& (\bbone-\bbw_1)^\top \bar{\bbB}_2^+ \bbw_2= \textstyle \sum_{t=1}^{\bar{T}}[\bar{\bbB}_2^{+\top}(\bbone-\bbw_1)]_{t}[\bbw_2]_t. \notag
\end{align}
These expressions reveal that both cost terms are linear and separable across the entries of $\bbw_2$. This is used to show that the optimal solution to \eqref{E:SC_OPT_v1_subp2_c1e} is given by the next lemma.
\begin{mylemma}\label{Lemma:GreedyTriangleSelection}
	Let us define the triangle-score vector $\bbs_2\in\reals^{\bar{T}}$ as $[\bbs_2]_t=\alpha_2+\beta_2[\bar{\bbB}_2^\top \bar{\bbX}_1 \bar{\bbX}_1^\top\bar{\bbB}_2]_{tt} + \gamma [\bar{\bbB}_2^{+\top}(\bbone-\bbw_1)]_{t}$ and let $\pi_2$ be 
	the permutation function $\{1,...,\bar{T}\} \rightarrow \{1,...,\bar{T}\}$ that orders the elements of $\bbs_2$ in an ascending manner so that $[\bbs_2]_{\pi_2(t)}\leq [\bbs_2]_{\pi_2(t+1)}$. 
 Then, the optimal solution to  \eqref{E:SC_OPT_v1_subp2_c1e} is 
	\begin{eqnarray}\label{E:optimal_greedy_wTriang}
		[\bbw_2]_t
		=\left\{\begin{matrix}	
			1,\;\;&\text{if} \;t\in \{\pi_2(i)\}_{i=1}^{T^{min}}\\
			0,\;\;&\text{otherwise}.
		\end{matrix}\right.	
	\end{eqnarray} 
\end{mylemma}


\noindent \emph{B. Optimization of the edge selection vector:}
Here, we consider $\bbw_2$ and $\bar{\bbX}_1$ given and optimize $\bbw_1$. The resultant problem is  
\begin{alignat}{2}
	\!\!&\!\min_{\bbw_1} &&   \alpha_1\|\bbw_1\|_0+ \beta_1\tr(\bbX_0\bbX_0^\top\bar{\bbB}_1\diag(\bbw_1)\bar{\bbB}_1^\top) \nonumber \\ \!\!&\! && + \gamma (\bbone-\bbw_1)^\top \bar{\bbB}_2^+ \bbw_2  \label{E:SC_OPT_v1_subp1_obj} 
	 \\ 
	\!\!&\! \mathrm{\;\;s. \;t. } && \bbw_1 \in \{0,1\}^{\bar{E}},  \; [\bbw_1]_{l}=1 \;\; \forall \;l\in \ccalEobs, \|\bbw_1\|_0 \geq E^{\min} . \nonumber 
\end{alignat} 

As before, we write the second and third terms as
\begin{align}
	&\tr(\bbX_0\bbX_0^\top\bar{\bbB}_1\diag(\bbw_1)\bar{\bbB}_1^\top)
	\!=\! \textstyle \sum_{l=1}^{\bar{E}}[\bar{\bbB}_1^\top\bbX_0\bbX_0^\top\bar{\bbB}_1]_{ll}[\bbw_1]_l,  \notag \\
	&(\bbone-\bbw_1)^\top \bar{\bbB}_2^+ \bbw_2= \textstyle \sum_{l=1}^{\bar{E}}[\bar{\bbB}_2^+\bbw_2]_{l}-\sum_{l=1}^{\bar{E}}[\bar{\bbB}_2^+\bbw_2]_{l}[\bbw_1]_l. \notag
\end{align}

The main difference relative to the problem \eqref{E:SC_OPT_v1_subp2_c1e} is that here some of the links are known. As a result, the counterpart to Lemma 1 for the edge-selection vector is given next.
\begin{mylemma}\label{Lemma:GreedyEdgeSelection}
	Let us define the edge-score vector $\bbs_1\in\reals^{\bar{E}}$ as 
 $[\bbs_1]_l=0$ if $l \in \ccalEobs$ and
 $[\bbs_1]_l=\alpha_1+\beta_1[\bar{\bbB}_1^\top\bbX_0\bbX_0^\top\bar{\bbB}_1]_{ll} - \gamma [\bar{\bbB}_2^+\bbw_2]_{l}$ otherwise and let $\pi_1$ be 
the permutation function $\{1,...,\bar{E}\} \rightarrow \{1,...,\bar{E}\}$ that orders the elements of $\bbs_1$ in an ascending manner so that $[\bbs_1]_{\pi_1(l)}\leq [\bbs_1]_{\pi_1(l+1)}$. Then, the optimal solution to  \eqref{E:SC_OPT_v1_subp1_obj} is 
	\begin{eqnarray}
	[\bbw_1]_l
	=\left\{\begin{matrix}	
		1,\;\;&\text{if} \;l\in \{\pi_1(i)\}_{i=1}^{E^{\min}}\\
		0,\;\;&\text{otherwise}.
		\end{matrix}\right.	
	\end{eqnarray} 
\end{mylemma}
Once again, the result shows that the greedy activation of the edges (according to the score defined in the lemma) is optimal.

\vspace{.15cm}
\noindent \emph{C. Optimization of the edge signals:} Here we consider $\bbw_1$ and $\bbw_2$ given and optimize over $\bar{\bbX}_1$. The resultant problem is  
\vspace{-.1cm}
\begin{alignat}{2}
	\!\!&\!\min_{\bar{\bbX}_{1}} &&  \; \beta_2\tr(\bar{\bbX}_1 \bar{\bbX}_1^\top\bar{\bbB}_2\diag(\bbw_2)\bar{\bbB}_2^\top) + \eta\|\bar{\bbTheta} \bar{\bbX}_1-\bbX_1^\ccalO\|_F^2  \label{E:SC_OPT_v1_subp3_obj} 
\end{alignat} 
The problem is convex and differentiable, with the solution being provided in the next lemma.
\vspace{-0.25cm}
\begin{mylemma}\label{Lemma:EdgeSignalsInterp}
\vspace{-0.105cm} The interpolated edge signals are given by
\begin{equation}
	\bar{\bbX}_1=  (\beta_2\bar{\bbB}_2\diag(\bbw_2)\bar{\bbB}_2^\top+\eta \bar{\bbTheta}^\top \bar{\bbTheta})^{\dagger} \eta \bar{\bbTheta}^\top\bbX_1^\ccalO.
\end{equation}
\end{mylemma}
This solution sets to zero all the rows of $\bar{\bbX}_1$ associated with edge indices that are neither observed nor induced by $\bbw_2$. 

\vspace{.1cm}
\noindent \emph{D. Block-coordinate SC-learning greedy algorithm:} 
Let $\ell=1,...,\ell^{\max}$ denote the iteration index and let $(\bbw_1^{(\ell)},\bbw_2^{(\ell)},\bar{\bbX}_1^{(\ell)})$ be the solution estimated at the end of iteration $\ell$. The steps to learn the topology of the SC from the smooth SC signals $\bbX_0$ and $\bbX_1^\ccalO$ are summarized in Fig.~\ref{F:exp_1}.a, which is labeled as ``GreedySCL''. 
Remarkably, the computational complexity of the algorithm is kept under control. Sorting the edges and triangles takes $O(\bar{E}\log(\bar{E}))$ and $O(\bar{T}\log(\bar{T}))$, respectively, although the complexity could be reduced by pruning some of the entries. Similarly, the computational complexity of the pseudoinverse can be reduced by exploiting the structure of the active triangles and observed edges and, if needed, by running an approximate solution based on gradient descent. 

	
	
		
		
		
		
	

\section{Numerical results}


Due to space limitations, we present here two illustrative examples and refer interested readers to the GitHub repository (\url{https://github.com/andreibuciulea/SC_Learning}) for full details on the simulation setup and additional experiments.

We employ Erdős-Rényi (ER) graphs with $N = 20$ nodes and edge probability $p = 0.4$.
Smooth node signals are generated from the true $\bbL_0$ with $P_0 = 100$.
For generating smooth edge signals, we consider that 50\% of the triangles are filled.
Subsequently, we generated edge signals with a small curl component across the filled triangles from $\bbL_{U} = \bbB_2\bbB_2^{\top}$ with $P_1 = 100$.
The evaluation metric used is the normalized squared Frobenius norm, defined as ${\rm NErr}(\bbL) = {\| \bbL^*-\hbL \|_F^2}/{\|\bbL^*\|_F^2}$ where $\hbL$ and $\bbL^*$ stand for the estimated and ground truth $\bbL$, respectively.
We evaluate our proposed method ``GreedySCL'' against ``SepSCL'' and ``RC''. SepSCL uses a greedy approach to estimate $\bbB_1$ \& $\bbB_2$ from $\bbX_0$ \& $\bbX_1$, without assuming any relationship between $\bbB_1$ \& $\bbB_2$. 
RC~\cite{zomorodian2010fast} estimates $\bbB_1$ \& $\bbB_2$ from the correlation of the node signals. 

\vspace{0.05cm}
\noindent\textbf{SC-learning from noisy node signals.}
We evaluate how effectively GreedySCL can use information from the edge signals to improve the estimation of both $\bbB_1$ \& $\bbB_2$. To achieve this, we assume that 80\% of the edge signals are known, and estimate $\bbL_0$ \& $\bbL_U$ while varying the standard deviation of the noise in the node signals.
Comparing the NErr for $\bbL_0$ between GreedySCL and SepSCL in Fig.~\ref{F:exp_1}.b, a significant difference in the error is observed. This discrepancy is mainly due to the fact that our model considers the presence of triangles for graph estimation, while SepSCL relies solely on node signals for graph estimation.
Conversely, RC is very close to SepSCL, suggesting that RC does not take advantage of the additional information to improve the estimation of $\bbL_0$.
Shifting the focus to $\bbL_U$, the error in $\bbL_U$ is significantly higher than in $\bbL_0$, underscoring the complexity of triangle estimation. 
However, at low noise levels, GreedySCL effectively uses the node signal information to reduce the estimation error associated with filled triangles, yielding better results than the alternatives.

\vspace{0.05cm}
\noindent\textbf{SC-learning from partially observed edge signals.} 
Here, we evaluate how well GreedySCL can utilize both the available node and edge data for joint estimation of $\bbB_1$ \& $\bbB_2$.
In Fig.~\ref{F:exp_1}.c, we observe for all methods, a decrease in NErr for $\bbL_0$ and $\bbL_U$ as the percentage of observed edge signals increases. The only exception is the $\bbL_0$ estimated by SepSCL, which arises because SepSCL estimates $\bbL_0$ without incorporating additional information from observed edges in $\bbX_1$. 
Comparing GreedySCL and SepSCL, significant benefits from joint estimation are apparent, particularly in improving the estimation of $\bbL_0$, with a smaller gain in the NErr of $\bbL_U$ for the same reason.
The RC method clearly does not perform well here.

\newpage

\bibliographystyle{IEEEtran.bst}
\bibliography{citations}
\end{document}